\documentstyle[aps,prb,epsf,floats]{revtex}

\begin{document}

\newcommand{\news}{{\tilde s}}
\newcommand{\newz}{{\tilde z}}
\newcommand{\scbo}{SrCu$_2$(BO$_3$)$_2$}

\twocolumn[\hsize\textwidth\columnwidth\hsize\csname@twocolumnfalse%
\endcsname

\title{Hole dynamics and photoemission in a $t$-$J$ model
       for SrCu$_2$(BO$_3$)$_2$}
\author{Matthias Vojta}
\address{
      Department of Physics,
      Yale University, P.O.Box 208120,
      New Haven, CT 06520-8120, USA
}
\date{\today}
\maketitle

\begin{abstract}
The motion of a single hole in a $t-J$ model for the two-dimensional spin-gap compound
SrCu$_2$(BO$_3$)$_2$ is investigated.
The undoped Heisenberg model for this system has an exact dimer eigenstate and shows a phase transition
between a dimerized and a N\'{e}el phase at a certain ratio of the
magnetic couplings.
We calculate the photoemission spectrum in the disordered phase using a generalized
spin-polaron picture.
By varying the inter-dimer hopping parameters we find a cross-over between
a narrow quasiparticle band regime known from other strongly correlated systems
and free-fermion behavior.
The hole motion in the N\'{e}el-ordered phase is also briefly considered.
\end{abstract}
]



Since the discovery of high-temperature superconductivity,
doped antiferromagnets (AF) have been studied intensively.
The pseudo-gap behavior observed in the high-$T_c$ cuprates
has stimulated great interest in systems with spin gaps.
Several new one- and two-dimensional spin gap systems have been found
experimentally. These materials are characterized by a disordered singlet ground
state and a finite gap to all spin excitations.
Some of the compounds
which have two-dimensional (2d) character include the coupled spin
ladder systems, SrCu$_2$O$_3$\cite{srcu2o3}, CaV$_2$O$_5$\cite{cav2o5},
(VO$_2$)P$_2$O$_7$\cite{vopo}, Cu$_2$(C$_5$H$_{12}$N$_2$)$_2$Cl$_4$\cite{cuHpCl},
and the plaquette RVB system, CaV$_4$O$_9$\cite{cav4o9}.

\begin{figure}
\epsfxsize=6.8 truecm
\centerline{\epsffile{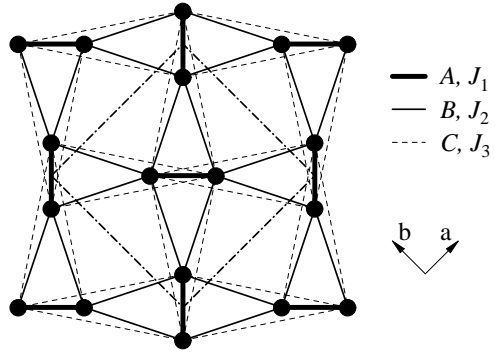}}
\caption{
Lattice structure of the Cu spins in \scbo,
with the three different exchange couplings and
the crystallographic axes a,b.
The dash-dotted lines denote a unit cell.
}
\label{FIG_LATTICE}
\end{figure}

Recently the two-dimensional spin gap system {\scbo} has been found
by Kageyama {\it et al.}\cite{kag98}.
It has a spin-singlet ground state with a spin gap $\sim 30$ K.
The substance has additional interesting features, e.g.,
the high-field magnetization was observed to have two plateaus at
1/4 and 1/8 of the full moment.
Recent work \cite{miy98} suggests that the underlying physics can be
understood on the basis of a two-dimensional $S={1\over 2}$ Heisenberg
model with antiferromagnetic nearest-neighbor ($J_1$, on links $A$) and
next-nearest-neighbor ($J_2$, on links $B$) couplings
on the lattice shown in Fig. \ref{FIG_LATTICE}.
The nearest-neighbor bonds $A$ define a unique singlet covering of the
lattice.
The Heisenberg model corresponding to Fig. \ref{FIG_LATTICE} (with $J_1$, $J_2$) is in fact
topologically equivalent
to the model considered by Shastry and Sutherland \cite{shastry}.
For this model the singlet product state forms an exact eigenstate of the
Hamiltonian at all couplings, and is the ground-state in a region where the
nearest-neighbor coupling $J_1$ dominates.
On the other hand, for $J_1\rightarrow 0$ the system becomes equivalent to the 2d square
lattice AF (with nearest-neighbor coupling $J_2$) which has a N\'{e}el-ordered
ground state.
The Shastry-Sutherland model can be complemented by a coupling $J_3$
on the links $C$; the singlet product state is an eigenstate of
this generalized model, too.
For $J_2=J_3$ the total spin on each $A$ bond is conserved, i.e., there is a macroscopic number
of conserved quantities, and each eigenstate is characterized by the number and
positions of triplets.

The Heisenberg model of Fig. 1 has been studied in Refs. \onlinecite {miy98,almi96,weih98,mh99}
using exact diagonalization, Schwinger boson and series expansion methods.
For $J_2=J_3$ it can be mapped onto a spin-1 model which shows a first order transition
between the singlet state and a N\'eel-ordered state at $J_2/J_1 = 0.4296$.
For $J_2\neq J_3$ the situation is less clear. For $J_3=0$ the numerical
results provide evidence for a weak first order singlet-N\'eel transition
at $J_2/J_1=0.70\pm 0.01$.
The disordered {\scbo} compound lies probably close to the transition line to a N\'eel state,
which explains the unusual temperature dependence of magnetic properties.
A fit of the susceptibility obtained from the model (with $J_1$ and $J_2$ only)
to the experimental data\cite{miy98} leads to the estimates of $J_1=100$ K
and $J_2/J_1=0.68$.
However, the actual ratio of the couplings $J_3/J_2$ is not known.

To include the doping degree of freedom into the Shastry-Sutherland
model we consider a standard $t-J$ model on the {\scbo} lattice:
\begin{eqnarray}
\mathcal{H} &=&
    \sum_{(i,j)\in A} \left (
 -  t_1 \: ( \hat{c}_{i\sigma}^\dagger \hat{c}_{j\sigma}^{} + h.c. )
 +  J_1 \: {\bf S}_i\cdot {\bf S}_j \right )
\nonumber\\
&+& \sum_{(i,k)\in B} \left (
 -  t_2 \: ( \hat{c}_{i\sigma}^\dagger \hat{c}_{k\sigma}^{} + h.c. )
  + J_2 \: {\bf S}_i\cdot {\bf S}_k \right )
\nonumber\\
&+& \sum_{(i,l)\in C} \left (
  - t_3 \: ( \hat{c}_{i\sigma}^\dagger \hat{c}_{l\sigma}^{} + h.c. )
  + J_3 \: {\bf S}_i\cdot {\bf S}_l \right )
\:.
\label{H}
\end{eqnarray}
The electron operators $\hat c_{i\sigma}^{\dagger}$ exclude double occupancies.
We have included hopping $t_3$ and interaction $J_3$ along the $C$ bonds.
If the $t-J$ model is derived as strong-coupling limit of a Hubbard model on the {\scbo}
lattice with on-site repulsion $U$, the ratio of the parameters is given by
\begin{equation}
{t_1^2 \over J_1} \:=\: {t_2^2 \over J_2} \:=\: {t_3^2 \over J_3} \:=\: {U \over 4}
\,.
\label{HUBBARD_REL}
\end{equation}

Although the compound {\scbo} has to our knowledge not been doped so far,
the study of the hole dynamics in this environment is an interesting and
challenging question.
Furthermore it may be possible that finite doping leads to the formation of
hole pairs and eventually to superconductivity.
In this paper we shall discuss the dynamics of a single hole in an otherwise
half-filled system using a generalized spin-polaron picture.
The one-hole spectral function for this case corresponds directly to the
result of an angle-resolved photoemission experiment on the undoped
compound which may give important information on the electronic
correlations and the exchange constants in \scbo.
We will show that the ratio $t_3/t_2$ tunes a cross-over 
between a narrow-band quasiparticle behavior and a regime where a free-fermion
peak dominates the spectrum.
The narrow-band behavior found here is similar to other 2d AF systems
\cite{Dagotto94}, it originates from the motion of a hole dressed with
spin fluctuations \cite{Dagotto94,Strings,MaHo,Riera97}.

To investigate the hole motion we consider a one-particle Green's function
describing the creation of a single hole with momentum $\bf k$
at zero temperature:
\begin{equation}
G({\bf k},\omega) \;=\; \sum_\sigma
  \langle\psi_0^N| {\hat c}_{{\bf k}\sigma}^\dagger {1 \over {z-H}}
               {\hat c}_{{\bf k}\sigma}
  |\psi_0^N\rangle
\label{HOLE_GF}
\end{equation}
where $z$ is the complex frequency variable, $z=\omega+i\eta$,
$\eta\rightarrow 0$.
$|\psi_0^N\rangle$ is the ground state of undoped system, i.e., with $N$ electrons on $N$
lattice sites.
%



To describe the dimerized phase we employ a bond operator representation \cite{SaBha,Gopalan} for
the spins on the nearest-neighbor bonds.
For each $A$ bond containing two $S=1/2$ spins we introduce bosonic operators for creation
of a singlet and three triplet states out of the vacuum $|0\rangle$:
$s^{\dagger}|0\rangle =
\frac{1}{\sqrt{2}}(|\uparrow\downarrow\rangle - |\downarrow \uparrow\rangle)$,
$t_{x}^{\dagger}|0\rangle =
  \frac{-1}{\sqrt{2}} (|\uparrow \uparrow\rangle - |\downarrow \downarrow\rangle)$,
$t_{y}^{\dagger}|0\rangle =
  \frac{i}{\sqrt{2}}(|\uparrow \uparrow\rangle + |\downarrow \downarrow\rangle)$,
$t_{z}^{\dagger}|0\rangle =
  \frac{1}{\sqrt{2}}(|\uparrow \downarrow\rangle + |\downarrow \uparrow\rangle)$,
where the constraint
$ s^{\dagger} s^{} + \sum_\alpha t_{\alpha}^{\dagger}t_{\alpha}^{} = 1$
has to be imposed on each bond to restrict the possible states to the
physical Hilbert space.
The original spins are related to the new boson basis operators by
$S_{1,2}^{\alpha} = \frac{1}{2} ( \pm s^{\dagger}  t_{\alpha}^{} \pm
 t_{\alpha}^{\dagger} s^{}  - i \epsilon_{\alpha\beta\gamma} t_{\beta}^{\dagger}
 t_{\gamma})$.
The Hamiltonian of the Shastry-Sutherland model written in terms of
the bond operators contains no terms which create triplet excitations from a
state containing only singlets which means
that the singlet product state
$|\phi_0\rangle = \prod_i s_i^\dagger |0\rangle$
is an exact eigenstate of the undoped system at all couplings.



If we remove one electron from an $A$ bond, a single-hole state on this
bond is created.
We introduce fermionic operators for bonding (symmetric) and antibonding (antisymmetric)
states of one electron (or hole) on an $A$ bond:
\begin{eqnarray}
a_{s,\sigma}^{\dag} |0\rangle =
\frac{1}{\sqrt{2}} ( {\hat c}^{\dag}_{1,\sigma} + {\hat c}^{\dag}_{2,\sigma}) |0\rangle \:,
\nonumber\\
a_{a,\sigma}^{\dag} |0\rangle =
\frac{1}{\sqrt{2}} ( {\hat c}^{\dag}_{1,\sigma} - {\hat c}^{\dag}_{2,\sigma}) |0\rangle \: .
\end{eqnarray}
The operators ${\hat c}^{\dag}_{1,2,\sigma}$ create an electron with spin $\sigma$ on one of the
two sites of an $A$ bond.
Hopping via $t_1$ leads to an `on-bond' energy of $\pm t_1$ for $a_a$ and $a_s$, respectively.
The other hopping terms in $\mathcal{H}$ permit the hole to hop between bonds;
the exchange interaction gives rise to spin fluctuations on
neighboring bonds.
The full interaction Hamiltonian in terms of $s$, $t$, and $a$ operators can be easily
found by calculating all possible one-hole matrix elements of the initial Hamiltonian $\mathcal{H}$
(\ref{H}) (see e.g. Ref. \onlinecite{Eder98}).

The evaluation of the Green's function (\ref{HOLE_GF}) is done using the
Mori-Zwanzig projection technique.
The set of dynamic variables is constructed from generalized path operators
~\cite{Strings,Vojta98}
which here create strings of triplet excitations attached to the hole.
Details of the calculational procedure can be found in Refs.
\onlinecite{Vojta98,Vojta99}.
In the disordered phase of the Shastry-Sutherland model the evaluation of the
matrix elements is simplified by the fact that the undoped ground state does
not contain background spin fluctuations, i.e.,
no cumulant expectation values are involved (cf. Ref. \onlinecite{Vojta99}).
In the present calculations we have employed up to 1800 dynamic variables with
a maximum path length of 3.
The neglect of the self-energy terms leads to a discrete set of poles for
the Green's functions, so the present approach cannot account for
linewidths.
In all figures we have introduced an artificial linewidth of $0.2 t_1$
to plot the spectra.



Before turning to the hole dynamics it is worth mentioning that
even a static vacancy (equivalent to the limit $t\rightarrow 0$)
has non-trivial consequences:
The pure singlet state with one spin removed, ${\hat c}_{i\sigma}|\phi_0\rangle$,
is no longer an eigenstate of $\mathcal{H}$.
The unpaired spin leads to triplet excitations in its neighborhood (`screening cloud'),
these are spatially confined as long as the spin gap is non-zero.

Now we consider the case of non-zero hopping.
We start with the `symmetric' choice $t_2 = t_3$, $J_2 = J_3$,
which allows to obtain several results analytically.
$J_2=J_3$ implies that triplets are strictly localized in the absence of the hole,
so any triplet created by hole hopping remains on its bond until the hole
returns.
$t_2=t_3$ reduces the number of possible inter-bond hopping processes.
The only non-zero hopping matrix elements are
$|\langle a_a s|{\mathcal H} | s a_a \rangle| =
 |\langle a_s t_\alpha |{\mathcal H} | t_\alpha a_s \rangle| =
 |\langle a_s s|{\mathcal H} | t_\alpha a_a \rangle| = t_2$
where $|XY\rangle = X_{i}^\dagger Y_{j}^\dagger |0\rangle$ is
a short-hand notation for a state of two neighboring bonds $i$ and $j$.
It follows that a hole in the antisymmetric state $a_a$ can freely propagate in a singlet
background without emission of triplet fluctuations (direct hopping), i.e.,
$|\psi_{a,\bf k}\rangle = \sum_i \exp({\rm i} {\bf k R}_i) a_{a,i}^\dag s_i |\phi_0\rangle$
is an exact eigenstate of $\mathcal{H}$.
In contrast, a hole being in the symmetric state $a_s$ always creates a triplet (and
converts into $a_a$) when it hops to a neighboring singlet bond.
This in turns means that any one-hole eigenstate of $\mathcal{H}$ which contains
components with $a_s$ also involves triplet excitations.
Since the triplets are localized and can only be created/removed by the hole,
the state itself is localized, and contributions from such eigenstates to the spectrum
are momentum-independent (non-dispersive).

\begin{figure}[!ht]
\epsfxsize=8.7 truecm
\centerline{\epsffile{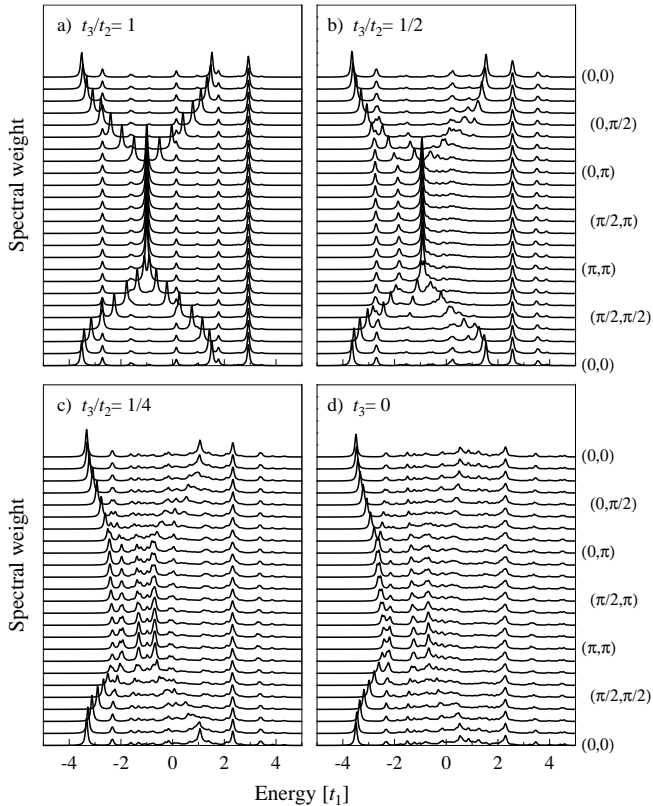}}
\caption{
Evolution of the one-hole spectral function $-{\rm Im}\: G({\bf k},\omega)$
under a change of the hopping ratio $t_3/t_2$.
a) `Symmetric' case $t_3/t_2=1$, $J_2/J_1=0.4$,
b) $t_3/t_2= 1/2$, $J_2/J_1=0.5$,
c) $t_3/t_2= 1/4$, $J_2/J_1=0.6$
d) $t_3=0$, $J_2/J_1=0.68$.
The other parameters are chosen according to eq. (\protect\ref{HUBBARD_REL})
with $U/t_1=4$.
The energies are measured in units of $t_1$ relative to the energy of
a localized hole.
The ratio of the magnetic couplings is chosen to place the system close
to the boundary of the singlet phase \cite{miy98,mh99}, i.e., the
spin gap in units of $J_1$ is nearly equal in the four cases.
}
\label{FIG_SPEC_CROSS}
\end{figure}

The calculated spectrum (Fig. \ref{FIG_SPEC_CROSS}a) is therefore easily understood:
It shows two dispersing bands which correspond to hole hopping in the antisymmetric
state $a_a$ with effective hopping amplitude $t_2=t_3$ through the square lattice of
rungs (two bands arise from the fact that the unit cell contains two rungs).
All other contributions are localized and involve symmetric hole states.
The corresponding wave functions can be modeled by a particle moving in an attractive potential centered
at a single site $i$.
A reasonable approximation for the states contributing to the spectrum is
$(u a_{s,i}^\dag s_i + \sum_{{\bf R},\alpha} v_{\bf R} a_{a,i+{\bf R}}^\dag s_{i+{\bf R}} t_{i\alpha}^\dag s_i)
|\phi_0\rangle$
where $u,v$ are some coefficients (typically $|u|\gg |v_{\bf R}|$ and $v_{\bf R}$ rapidly decaying
with distance).
Components with more than one triplet have coefficients much smaller than the one-triplet
coefficients $v_{\bf R}$.
Variation of the model parameters (keeping $t_2 = t_3$, $J_2 = J_3$) only results in small changes
in the spectrum of Fig. \ref{FIG_SPEC_CROSS}a since the dominating bands are determined by $t_2$
only.
Note, however, that varying $t_1/t_2$ shifts the dispersing bands w.r.t. to the localized peaks,
i.e., it can induce a level crossing at momentum $(0,0)$,
so $t_1\gg t_2$ eventually leads to a localized one-hole ground state.

Having understood the special situation at $t_2 = t_3$, $J_2 = J_3$, we turn to
the general case. We assume $J_3 < J_2$ [and $t_3 < t_2$ because of (\ref{HUBBARD_REL}) ]
since the model behavior is symmetric with respect to the interchange of the $B$ and $C$
bonds.
In the following we discuss the $t$ and $J$ parameters separately, but we keep in mind that
they are usually connected by the relation (\ref{HUBBARD_REL}).
For $J_2 \neq J_3$ the triplets are no longer completely localized, however,
as is known from the undoped model, triplet hopping does not occur up to 6th order
of perturbation theory in $J_2$, and so the triplet dispersion is weak.
Consequently, $J_2 \neq J_3$  introduces a weak dispersion into the `background features'
of the spectrum.
The effect of $t_2 \neq t_3$ on the spectrum is more pronounced:
Hopping of the antisymmetric hole state now also emits triplets, i.e.,
$|\psi_{a,\bf k}\rangle$ is no longer an exact eigenstate of $\mathcal{H}$.
Therefore the well-defined dispersing bands of Fig. \ref{FIG_SPEC_CROSS}a mix with the
background features.
Spectral weight is transferred to the bottom of the spectrum which
is easily understood from the fact that any triplet excitation increases the
energy by at least the gap size.
Figs. \ref{FIG_SPEC_CROSS}a - \ref{FIG_SPEC_CROSS}d show the evolution of the spectrum
when going from $t_3/t_2=1$ to $t_3/t_2=0$;
we have fixed $U/t_1=4$ and chosen the remaining parameters such that the size of the
spin gap is (approximately) preserved.
For $t_3=J_3=0$, Fig. \ref{FIG_SPEC_CROSS}d, we arrive at a situation with a narrow band
at the bottom of the spectrum with dispersion minimum at $(0,0)$;
this band corresponds to the motion of a hole surrounded by triplet fluctuations
(spin polaron).
At higher energies a weak background is visible, it arises from excited polaron
states.
In addition, we note that smaller relative hopping strengths $t/J$
suppress triplet fluctuations and reduce the high energy background,
whereas larger values of $t/J$ lead to an incoherent spectrum since the size
of the spin polaron increases.

\begin{figure}
\epsfxsize=7.5 truecm
\centerline{\epsffile{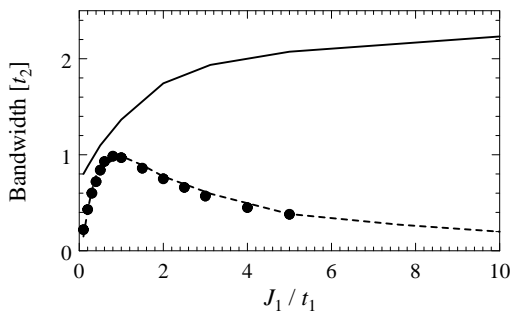}}
\caption{
Comparison of hole QP bandwidths in units of the (inter-dimer) in-plane hopping
amplitude.
Solid: {\scbo} lattice, $t_2/t_1=0.82$, $J_2/J_1=0.68$, $t_3=J_3=0$, present calculation.
Dashed: Square lattice, obtained by the methods used in this paper.
Circles: Square lattice SCBA results from Ref. \protect\onlinecite{MaHo}.
}
\label{FIG_BANDW1}
\end{figure}

To make contact with possible experiments we examine briefly the properties
of the pronounced bands at the lower edge of the spectrum which would be
visible in a photoemission spectrum.
The experimentally measured bandwidth can be used to determine
the ratio $t/J$, or equivalently, the on-site repulsion $U$.
For the (experimentally unrealistic) case of $t_2=t_3$, $J_2=J_3$,
two bands should be observed as in Fig. \ref{FIG_SPEC_CROSS}a.
The width of each of the bands is given by $4 t_2$ since the dispersion is
the one of a free fermion.
More likely, the material has $t_3 \ll t_2$, $J_3 \ll J_2$, which corresponds
to Fig. \ref{FIG_SPEC_CROSS}c or d.
As a guide we plot in Fig. \ref{FIG_BANDW1} the bandwidth for $t_3=J_3=0$ as
function of $J_1/t_1$.
In contrary to the well-known square-lattice case (shown for comparison)
the bandwidth is finite for $J\rightarrow 0$.
The reason is the possibility of direct hopping (without emission of spin excitations);
this is not the case in a square lattice antiferromagnetic background where
hopping always creates spin defects which have to be removed by exchange
processes.
For $t/J\rightarrow 0$ the bandwidth saturates (in units of $t_2$) at a non-trivial
value which arises from the finite cloud of triplet excitations around a
static hole.
(Of course, for $J_2=J_3=0$ this effect is absent, and the bandwidth becomes
$2 t_2$ in the limit of $t/J\rightarrow 0$.)

We have also calculated the spin correlations near the mobile hole.
Except for the case corresponding to Fig. \ref{FIG_SPEC_CROSS}a which shows
no inter-dimer correlations in the ground state,
the hole always introduces antiferromagnetic correlations between different
dimers in its vicinity.
However, for the possibly relevant values ($J_2/J_1=0.68$, $U/t_1\sim 2-10$)
the dimerization is strong and the polaron size can be estimated to be
smaller than 3 lattice spacings.

We briefly mention that the present analysis can be extended to the N\'{e}el-ordered
phase of the Shastry-Sutherland model as well.
Therefore a condensate of one type of triplets is introduced by a proper
transformation of the basis states on each bond \cite{Vojta99};
the condensation amplitude can be extracted from series expansion results
\cite{weih98}.
The ground state of the undoped system is obtained by an expansion around a
modified product state (which is the N\'{e}el state in the case of dominating
$J_2$); here of cause the ground state contains fluctuations around the
product state.
The spin polaron consists of spin deviations from the undoped background
state in the vicinity of the hole.
The results show that with increasing antiferromagnetic correlations
the band minimum is shifted from $(0,0)$ to $(\pi,\pi)$ (this is equivalent
to $(\pi/2,\pi/2)$ in the Brillouin zone of the square lattice defined by the
$B$ bonds);
the bands approach the shape known from the square-lattice
antiferromagnet.



Summarizing, in this paper we have studied the
one-hole dynamics in a $t-J$ model for the 2d spin gap
material \scbo.
Using a generalized spin-polaron concept together with an expansion around
a singlet product state we have calculated the one-hole spectral
function.
It shows an interesting cross-over from free-fermion
behavior to correlated behavior under a variation of the
ratio $t_2/t_3$ of 2nd and 3rd-nearest-neighbor coupling.
At $t_3=t_2$ the main contributions to the spectrum can be described by the
hopping of the bare hole between
the singlet rungs with a tight-binding dispersion.
In contrast, for $t_3\ll t_2$ one finds a narrow band at the bottom of the
spectrum and a background at higher energies;
this structure can be attributed to the motion of a hole dressed with spin fluctuations
(similar to other strongly correlated systems).
We add that very large $t/J$ leads to one-hole ground states with
higher spin and ferromagnetic correlations in spirit of the Nagaoka effect,
this has not been considered here.
The hole dynamics at finite doping and the possibility of
hole pairing are interesting subjects of future research.


Financial support from the DFG (VO 794/1-1) is gratefully acknowledged.



\end{document}